\documentclass[prd,aps,showpacs,tightenlines,10pt, reprint,nofootinbib,superscriptaddress,floatfix,twocolumn]{revtex4-2}

\usepackage{graphicx}% Include figure files
\usepackage{dcolumn}% Align table columns on decimal point
\usepackage{bm}% bold math
\usepackage{amsmath,amssymb}
\usepackage{xcolor}
\usepackage{ytableau}
\usepackage[colorlinks,linkcolor=blue]{hyperref}

\usepackage{orcidlink}

\usepackage{academicons}
\definecolor{orcidlogocol}{HTML}{A6CE39}
\newcommand{\orcid}[1]{\href{https://orcid.org/#1}{\textcolor[HTML]{A6CE39}{\aiOrcid}}}

\allowdisplaybreaks

\begin{document}

\preprint{XYZ}

\title{Reexamining aspects of spacetime-symmetry breaking with CMB polarization}

\author{Nils A. Nilsson\,\orcidlink{0000-0001-6949-3956}}
\email{nils.nilsson@obspm.fr}
\author{Christophe Le Poncin-Lafitte\,\orcidlink{0000-0002-3811-1828}}%
\email{christophe.leponcin@obspm.fr}
\affiliation{SYRTE, Observatoire de Paris, Université PSL, CNRS, LNE, Sorbonne Universit\'e, 61 avenue de l’Observatoire, 75 014 Paris, France}%

\date{\today}% It is always \today, today,

\begin{abstract}
The linear polarization of the Cosmic Microwave Background (CMB) is highly sensitive to parity-violating physics at the surface of last scattering, which might cause mixing of E and B modes, an effect known as {\it cosmic birefringence}. This has until recently been problematic to detect due to its degeneracy with the instrument polarization miscalibration angle. Recently, a possible detection of a non-zero cosmic-birefringence angle was reported at $\beta={0.35^\circ}\pm0.14^\circ$, where the miscalibration angle was simultaneously determined and subtracted from the analysis. Starting from this claim, we exploit a simple map of $\beta$ to the coupling constant of a parity-violating term in a generic effective-field theory for Lorentz and CPT violation. We show that the reported constraint on $\beta$ is consistent with current one-sided upper bounds from CMB studies of spacetime-symmetry breaking, and we discuss the implications and interpretation of this detection.
\end{abstract}

\maketitle

%\tableofcontents

\section{Introduction}
Measurements of the temperature anisotropies and polarization patterns of the Cosmic Microwave Background (CMB) have provided detailed information about the state of the Universe at very high redshift ($z\approx 1100$), and has been instrumental in establishing the Lambda Cold Dark Matter ($\Lambda$CDM) cosmological model. The CMB polarization is especially sensitive to parity-violating physics\footnote{or \emph{any} modifications to standard electrodynamics}, which until recently was only known to be present in weak interaction. In the standard model, the CMB acquires linear polarization due to Compton scattering during reionization, which can be affected by new physics in the early Universe or by non-standard photon propagation.

Recently, hints of cosmic birefringence, the rotation of E modes into B modes in the CMB, were discussed in a series of papers by Komatsu et al~\cite{Minami:2020odp,Komatsu:2022nvu,Nakatsuka:2022epj,Eskilt:2022cff}, and at this point, the statistical significance for a non-zero birefringence angle $\beta$ has reached $3.6\sigma$ \cite{Eskilt:2022cff}, with no evidence of frequency dependence \cite{Eskilt:2022wav}. This mixing of CMB parity eigenstates is consistent with the production through an axion-like field (see Section~\ref{sec:axph}) and disfavours EB production through Faraday rotation~\cite{Eskilt:2022cff,Scoccola:2004ke}. Even if B modes, either primordial or generated through weak lensing, are present in the CMB, a parity-violating term is still necessary to generate the non-zero mixing angle in order for cosmic birefringence to appear. Several theoretical mechanisms are available for this purpose, for example coupling of the electromagnetic field with an axion-like particle (ALP) \cite{Finelli:2008jv}, Chern-Simons interactions \cite{Carroll:1989vb,Carroll:1998zi,Li:2008tma}, and certain kinds of quintessence fields \cite{Giovannini:2004pf,Balaji:2003sw}.

As the most redshifted electromagnetic radiation available, the CMB is a sensitive probe of early-Universe physics, and in order to capture the parity-odd nature of cosmic birefringence, a non-standard term needs to be added to the Lagrangian, where the axion-photon interaction can be readily represented by a {\it CPT}-odd term.
In the photon sector, such symmetry violations have been studied extensively using a generic effective-field theory framework \cite{Colladay:1996iz,Colladay:1998fq} as a basis for experimental searches; studies using gamma-ray bursts \cite{Kostelecky:2013rv,Toma:2012xa,Laurent:2011he,Wei:2019nhm}, spectropolarimetry of cosmological sources \cite{Kostelecky:2001mb}, rotating optical resonators \cite{Zhang:2021sbx,Chen:2016eli}, optical ring cavities \cite{Michimura:2013via}, and many more have yielded tight constraints on Lorentz and $CPT$ symmetry, and are yet to detect a non-zero signal; the tightest constraints to date have been obtained using CMB polarization \cite{Caloni:2022kwp}. For an extensive list of all available constraints (updated annually), the reader is referred to \cite{Kostelecky:2008ts}. 

When studying CMB polarization it is necessary to take the instrument miscalibration angle (often referred to as $\alpha$) into account, as any remaining $\alpha$ will be degenerate with $\beta$ and will inevitably mimic an isotropic birefringence signal as $\beta^{\prime} = \alpha+\beta$; for example, the {\it Planck} team has determined $\beta^\prime = 0.31\pm0.05$, not including systematic uncertainties \cite{Planck:2016soo}. Several different methods for determining $\alpha$ has recently been employed in the literature; in \cite{Minami:2020odp}, the authors used {\it Planck} Public Data Release 3 HFI maps to obtain a global fit on $\beta$ and three different constraints on $\alpha$ in three different frequency bands. Furthermore, in \cite{Eskilt:2022cff}, the authors refine the analysis in \cite{Minami:2020odp} to take into account polarized thermal dust emission in the galactic plane, which is the dominant foreground contribution to the polarization signal. Both of these approaches assume vanishing intrinsic EB correlation spectrum at the Last Scattering Surface (LSS), since this contribution lies beneath the sensitivity curves of current instruments\footnote{Such an effect may be taken into account, and in general shows a different dependence on the angular index $\ell$ \cite{Thorne:2017jft}}. 

The purpose of the present paper is to highlight an important implication of a potential confirmed detection of a non-zero cosmic-birefringence angle, namely that it is consistent with being generated through a Lorentz and $CPT$\footnote{See the theorem by Greenberg \cite{Greenberg:2002uu}.} breaking operator in the photon sector under certain assumptions, and that this result actually is consistent with existing (one-sided upper) constraints on spacetime-symmetry breaking. We organize the discussion as follows: In Section \ref{sec:EFT} we write down the photon sector including leading-order spacetime-symmetry breaking terms and show the resulting expressions for the EB power spectrum, which is parity-odd. In Section~\ref{sec:axph} we write down the corresponding expressions for when the parity-violating term arises from an axion-photon coupling. Here, we also discuss the measurement of cosmic birefringence, as well as the resulting constraints on spacetime-symmetry breaking, and we highlight the difference between the two. We discuss our results in Section~\ref{sec:disc}. We use natural units where $G = \hbar = 1$ throughout this paper.

\section{Effective-field theory}\label{sec:EFT}
In this section, we gather some useful results which are also found elsewhere, see for example \cite{Caloni:2022kwp} and \cite{Eskilt:2023nxm}. To arbitrary order in mass dimension $d$, we can write the extended Maxwell Lagrange density using effective-field theory (EFT) as \cite{Kostelecky:2001mb}
\begin{equation}
\begin{aligned}
    \mathcal{L} \sim -\tfrac{1}{4}F_{\mu\nu}F^{\mu\nu}+&\tfrac{1}{2}\epsilon^{\alpha\beta\mu\nu}A_\beta (\hat{k}_{\rm AF})_\alpha F_{\mu\nu} \\&- \tfrac{1}{4}(\hat{k}_{\rm F})^{\alpha\beta\mu\nu}F_{\alpha\beta}F_{\mu\nu},
    \end{aligned}
\end{equation}
where $\epsilon^{\alpha\beta\mu\nu} = E^{\alpha\beta\mu\nu}/\sqrt{-g}$ is the LeviCivita tensor, $A_\mu$ is the Maxwell 4-potential, and $F_{\mu\nu}=2\partial_{[\mu}A_{\nu]}$ the associated field strength. The first term generates conventional electrodynamics, and the other terms violate spacetime symmetries. The quantities $\hat{k}_{\rm F}$ and $\hat{k}_{\rm AF}$ are derivative operators defined as
\begin{equation}
    \begin{aligned}
    (\hat{k}_{\rm F})^{\kappa\lambda\mu\nu} &\equiv \sum_{d \,\in \,2\mathbb{Z}} (k_{\rm F}^{(d)})^{\kappa\lambda\mu\nu\alpha_1\hdots\alpha_{d-4}}\partial_{\alpha_1}\hdots \partial_{\alpha_4} \\
    (\hat{k}_{\rm AF})_{\kappa} &\equiv \sum_{d \,\in \,2\mathbb{Z}+1} (k_{\rm AF}^{(d)})_\kappa^{~\alpha_1\hdots\alpha_{d-3}}\partial_{\alpha_1}\hdots \partial_{\alpha_3},
    \end{aligned}
\end{equation}
where the coefficient $(k^{(d)}_{\rm AF})_\kappa$ is conformally invariant and $\it CPT$ odd, whereas $(k^{(d)}_{\rm F})^{\alpha\beta\mu\nu}$ is {\it CPT} even. These coefficients arise due to the symmetry breaking, and the associated Nambu-Goldstone modes could in principle play the role of the photon \cite{Bluhm:2004ep}. Since we are interested only in {\it CPT}-violating physics we will set $(k_{\rm F})^{\alpha\beta\mu\nu}$ to zero from now on; moreover, we will focus on the lowest-order operators, and we therefore truncate the series at $d=3$, at which point the coefficient $(k_{\rm AF}^{(3)})_\kappa$ appears with no associated partial derivatives. 

To make contact with the CMB, we note that there exists a known analogy between symmetry-breaking electrodynamics in vacuum and conventional electrodynamics in an anisotropic medium \cite{Kostelecky:2001mb,Gubitosi:2010dj,Caloni:2022kwp} through a modified Amp\`ere-Maxwell equation, which in our case reads
\begin{equation}
    \chi^{ij}=-2i\frac{c}{\omega}\epsilon_{ikj}(k_{\rm AF})^k-2i\left(\frac{c}{\omega}\right)^2(k_{\rm AF, 0})k^k,
\end{equation}
where $\chi^{ij}$ is the susceptibility tensor and $\omega$, $k$ are the comoving angular velocity and wavenumber. Using the formalism developed in \cite{Lembo:2020ufn} and used in \cite{Caloni:2022kwp}, the susceptibility tensor $\chi^{ij}$ can be reformulated as a mixing matrix between the Stokes parameters usually denoted $U$, $V$ and $Q$, where the combination of the linear-polarization parameters $Q(\hat{n})\pm iU(\hat{n})$ measured in the direction $\hat{n}$ transform as a spin-2 ($\pm 2$) object under parity inversion. Using spin-weighted spherical harmonics, this combination can be decomposed into the well-known E and B-mode polarizations as
\begin{equation}
    Q(\hat{n})\pm iU(\hat{n})=-\sum_{\ell m}\left(E_{\ell m}\pm iB_{\ell m}\right) _{\pm 2}Y_{\ell m}(\hat{n}),
\end{equation}
where the polarizations transform with opposite signs under a parity transformation $\hat{n} \to -\hat{n}$ as $E_{\ell m} \to \left(-1\right)^\ell E_{\ell m}$ and $B_{\ell m} \to \left(-1\right)^{\ell+1}B_{\ell m}$. 
We define the angular power spectra as
\begin{equation}
    C_\ell^{\rm XX^\prime} \equiv \left(2\ell +1\right)^{-1}\sum_m X_{\ell m}X^{\prime*}_{\ell m},
\end{equation}
where $X,X^\prime= \{E,B\}$. From this definition it is evident that the $EE$ and $BB$ power spectra are {\it even} under parity inversion, whereas the $EB$ spectrum is {\it odd}, and thus signals parity violation.

Using techniques outlined in \cite{Murai:2022zur,Caloni:2022kwp}, we can relate the EB polarization spectrum to the EE and BB spectra as
\begin{equation}\label{eq:SMECMBcorrs}
\begin{aligned}
    C_\ell^{\rm EB, k} &= 4c\sqrt{((\bar{k}_{\rm AF})_0)^2}(\widetilde{C}_\ell^{\rm EE}-\widetilde{C}_\ell^{\rm BB}),\\
    \end{aligned}
\end{equation}
where a tilde denotes the quantity in the absence of the parity-odd interaction term. The bar over $\bar{k}_{\rm AF}$ indicates that the quantity has been averaged along the line-of-sight from the surface of last scattering to the present time as
\begin{equation}
    (\bar{k}_{\rm AF})_0 \equiv \int_{\eta_0}^{\eta_{\rm LSS}}(k_{\rm AF})_0 \, d\eta
\end{equation}
where $\eta$ denotes the conformal-time coordinate, and we have assumed that the CMB photons propagate on standard lightcones; in principle, there are higher-order corrections to the dispersion relations proportional to $k_{\rm AF}$ to the n$^{\rm th}$ power, but which we discard, knowing that they will be very small. Therefore, $C_\ell^{\rm EB, k}$ constitutes the parity-violation induced spectrum sourced by the symmetry-breaking term $({\bar{k}_{\rm AF}})_0$, which is exactly a signal of cosmic birefringence. As was pointed out in \cite{Lue:1998mq}, a non-zero value of the parity-odd power spectra would indicate a preferred direction (and hence spacetime-symmetry breaking) present in the Universe. 

In order to take advantage of existing constraints on the effective-field theory coefficients $(\bar{k}_{\rm AF})_0$ \cite{Kostelecky:2008ts}, the prefactor in Eq.~\eqref{eq:SMECMBcorrs} can be rewritten as
\begin{equation}
    \gamma = 16c^2((\bar{k}_{\rm AF})_0)^2,
\end{equation}
which can be directly mapped to a standard coefficient by using the Stokes parameters and expanding in spherical harmonics, after which it was found in \cite{Caloni:2022kwp} that
\begin{equation}\label{eq:kV3}
    |k^{(3)}_{(V),00}|= \sqrt{\frac{\pi}{4c^2}}\frac{1}{\eta_0-\eta_{\rm LSS}} \sqrt{\gamma},
\end{equation}
where $c$ is the speed of light and $\eta_0$, $\eta_{\rm LSS}$ is the conformal time today and at the last scattering surface, respectively. This coefficient is related to the cosmic-birefringence angle and is in principle degenerate with the instrument miscalibration $\alpha$. It is thanks to the disentangling of these two angles carried out in \cite{Minami:2020odp} and others that we are able to map the obtained constraints on cosmic birefringence to the spacetime-symmetry breaking coefficients $k_{\rm AF}$.

By using the best-fit values from $Planck$ 2018 (TT+TE+EE+lowE)~\cite{Planck:2018vyg}, we can write
\begin{eqnarray}\label{eq:etaint}
    c(\eta_0-\eta_{\rm LSS}) = \frac{c}{H_0}\int\frac{dz}{E(z)} \approx 9444 \text{ Mpc},
\end{eqnarray}
in terms of redshift $z$, where $E(z)=\sqrt{\Omega_m^0(1+z)^3+\Omega_r^0(1+z)^4+\Omega_\Lambda^0}$ is the flat $\Lambda$CDM Hubble function, and Eq.~\eqref{eq:kV3} now reads
\begin{equation}
    |k^{(3)}_{(V),00}| = \left(6\cdot 10^{-43} \text{ GeV}\right) \sqrt{\gamma},
\end{equation}
where we now have a relation between the measured amplitude of the EB power spectrum (isotropic cosmic birefringence) and the existing constraints on $|k^{(3)}_{(V),00}|$. In \cite{Caloni:2022kwp}, the most stringent constraints to date\footnote{See also \cite{Kostelecky:2008be} for older constraints.} were found using CMB polarization
\begin{equation}\label{eq:kv}
    \begin{aligned}
        |k^{(3)}_{(V),00}| <& 6.81 \cdot 10^{-44} \text{ GeV}, \quad \it Planck\\
        |k^{(3)}_{(V),00}| <& 1.54 \cdot 10^{-44} \text{ GeV}, \quad \begin{aligned}&{\it Planck}+\rm BCII\\&+\rm ACT\end{aligned}
    \end{aligned}
\end{equation}
where BCII and ACT denote Bicep II and the Atacama Cosmology Telescope, respectively. From this, we obtain
\begin{equation}\label{eq:gamma}
    \begin{aligned}
        \sqrt{\gamma} <& 11.35 \cdot 10^{-2}, \quad \it Planck\\
        \sqrt{\gamma} <& 2.57 \cdot 10^{-2}, \quad \begin{aligned}&{\it Planck}+\rm BCII\\&+\rm ACT,\end{aligned}
    \end{aligned}
\end{equation}
i.e. amplitude of the EB power spectrum is on the order of a few percent of the EE and BB spectra. 

When comparing constraints below, we stick to constraints obtained through CMB only, specifically with {\it Planck}, even though stronger constraints can be obtained when combining probes (see Eqs.~(\ref{eq:kv})-(\ref{eq:gamma})); the reasons are twofold: first, local experiments need not be averaged over cosmic history, which is in essence a smearing of the different coefficients; second, using multiple CMB experiments necessitates a more careful treatment of the miscalibration angles.

\section{Axion-photon coupling}\label{sec:axph}
Having seen that a {\it CPT}-breaking effective-field theory term may generate non-zero EB correlation, we turn to a specific model, a candidate mechanism for generating cosmic birefringence consistent with the current detection; a Chern-Simons type coupling between the Standard Model photon and an axion-like particle (ALP), with the interaction term~\cite{Carroll:1989vb,Ferreira:2020fam,Harari:1992ea}
\begin{equation}
    \mathcal{L}_{\rm int} \sim \tfrac{1}{4}g_{\phi\gamma}\phi F_{\mu\nu}\widetilde{F}^{\mu\nu},
\end{equation}
where $\phi$ is the axion-like field, $g_{\phi\gamma}$ is the coupling constant, and 
\begin{equation}
\widetilde{F}^{\mu\nu} = \frac{\epsilon^{\mu\nu\rho\sigma}}{2\sqrt{-g}}F_{\rho\sigma}
\end{equation}
is the Maxwell dual. Such a coupling to a time-dependent ALP ($\phi = \phi(t)$) rotates the plane of linear polarization for photons (without influencing the Einstein equations) and gives rise to isotropic cosmic birefringence with the angle
\begin{equation}
    \beta = \tfrac{1}{2}g_{\phi\gamma}\int^{\eta_{\rm LSS}}_{\eta_0}d\eta\frac{\phi^\prime}{a},
\end{equation}
where prime denotes a derivative w.r.t $\eta$, and $a$ is the cosmic scale factor. The observed EB spectrum generated by the ALP then reads \cite{Minami:2020odp}
\begin{equation}
    C_\ell^{\rm EB, \phi} = \tfrac{1}{2}\sin{4\beta}(\widetilde{C}_\ell^{\rm EE}-\widetilde{C}_\ell^{\rm BB}),
\end{equation}
and we are now in a position to compare predictions from effective-field theory in Eq.~(\ref{eq:SMECMBcorrs}) and the axion-photon coupling above. The angle $\beta$ is (as for the coefficient $(\bar{k}_{\rm AF})_0$ in the previous section) degenerate with the instrument miscalibration angle $\alpha$.

We focus now on the reported non-zero birefringence angle ${\beta=0.35^\circ}\pm0.14^\circ \,(1\sigma)$ in \cite{Minami:2020odp}, which is non-zero at $2.4\sigma$; using this value for $\beta$, the EB spectrum amplitude in $C_\ell^{\rm EB, \phi}$ is\footnote{Disregarding the $2\pi n$ symmetry, since $\beta$ represents a small, positive anti-clockwise rotation.}
\begin{equation}\label{eq:klimit}
    \tfrac{1}{2}\sin{4\beta} = \left(1.22\pm0.49\right)\cdot 10^{-2}
\end{equation}
at $1\sigma$ confidence level. We note that this results was arrived at by disregarding possible pollution of the signal from the galactic foreground, which is possibly a large contribution of EB signal; however, it was shown in \cite{Eskilt:2022cff} that the signal is non-zero even if the galactic plane is masked. In \cite{Eskilt:2022cff}, the authors also present a constraint on $\beta$ which significantly smaller error bars, at $\beta=0.34\pm0.09$, corresponding to a non-zero detection at $3.6\sigma$. We choose to not use this constraint here, since the foreground contribution and method for disentangling $\beta$ from the miscalibration angle $\alpha$ makes it less straightforward to compare the constraints to those from effective-field theory found in \cite{Caloni:2022kwp}. We now use the opposite approach compared to Section~\ref{sec:EFT} and derive the implied bounds on $|k^{(3)}_{(V),00}|$ directly from the constraints on $\beta$. We have that 
\begin{equation}
|k^{(3)}_{(V),00}| = (6\cdot 10^{-43} \text{ GeV})(\tfrac{1}{2}\sin{4\beta}),
\end{equation}
from which the constraints read $|k^{(3)}_{(V),00}| = \left(7.32\pm2.94\right)\cdot 10^{-45} \text{ GeV}$ at $1\sigma$, which, since it is non-zero, suggests the presence of some unknown systematic\footnote{Although it should be noted here that we derived this from the non-zero measurement of $\beta$.} (as it appears to be a signal of spacetime-symmetry breaking), and since the error bars are algebraically mapped to $|k^{(3)}_{(V),00}|$, this is a non-zero signal at $2.4\sigma$;
we also see that this lies within the one sided upper limit found in  \cite{Kostelecky:2009zp,Carroll:1989vb,Mewes:2008gh,Caloni:2022kwp} even when using the strongest constraint from the {\it Planck}+BCII+ACT combination. For comparison purposes, we can rewrite the above bound using the $1\sigma$ upper limit to find $|k^{(3)}_{(V),00}|<1.026 \cdot 10^{-44}$ GeV, which can readily be seen to be stronger than the result (\ref{eq:kv}), which did not take into account the miscalibration angle. It is important to note that since the limit is on the {\it magnitude} of the coefficient $k_{(V),00}^{(3)}$, our constraint on $|k^{(3)}_{(V),00}|$ can be viewed as covering zero; however, together with the knowledge that $\beta \neq 0$ to high confidence, this implies that $|k^{(3)}_{(V),00}|\neq 0$ as well.

As a non-zero $|k^{(3)}_{\rm (V),00}|$ implies the rotation of E modes into B modes, EB modes will be present in the CMB through Eq.~\eqref{eq:SMECMBcorrs}. In the case of no lensing and no primordial B modes, the EB spectrum is simply a scaling of the Lorentz invariant EE spectrum\footnote{There will also be further leakage into the BB modes, which we do not show here, since the amplitude is very small.}. We note here also that we have neglected all other intrinsic sources of primordial EB correlations, in line with \cite{Minami:2020odp}. Such correlations could for example be induced in theories containing chiral primordial fluctuations, one example of with is Ho\v{r}ava-Lifshitz gravity \cite{Horava:2009uw}, as was shown in \cite{Takahashi:2009wc}. The possibility of sourcing isotropic cosmic birefringence from such primordial B modes was investigated in a model-independent manner in \cite{Fujita:2022qlk}, where it was found that there is a significant overproduction in the primordial BB spectrum which exceeds current limits from SPTPol and POLARBEAR \cite{Fujita:2022qlk}; therefore, this explanation can be said to be ruled out.

The situation would grow more complicated if we were to also consider anisotropic cosmic birefringence, which involves the spatial components of the coefficient $k_{\rm AF}$; these quantities shows a degeneracy with the temporal components. In principle, this could cancel out the non-zero signal we discuss in this letter, but this would require the presence of anisotropic cosmic birefringence, which has not yet been observed; a full analysis of these two competing effects lies beyond the scope of this work.

\section{Discussion \& Conclusions}\label{sec:disc}
In this paper, we have pointed out a parallel between the recent detection of isotropic cosmic birefringence and existing constraints of photon-sector spacetime-symmetry breaking. We point out that a non-zero cosmic-birefringence angle is actually consistent with photon-sector spacetime-symmetry breaking in the form of a Chern-Simons type $F\widetilde{F}$ term; moreover, we show that this is consistent with current constraints on spacetime-symmetry breaking from the CMB. Since the existing limits are one-sided upper bounds, it is possible to satisfy them and simultaneously measure a non-zero signal. 
We do not contend here that {\it CPT}-breaking modifications of electrodynamics exists in addition to a pseudoscalar axion, but we point out that a {\it CPT}-odd EFT coupling can mimic the EB correlations induced by an axion, and vice versa, and that the interpretation behind these degenerate mechanisms is completely different.

Constraints on $|k^{(3)}_{(V),00}|$ have also been obtained from probes other than CMB polarization, which besides being weaker than that of \cite{Caloni:2022kwp} can serve to break the degeneracy between the observed cosmic birefringence and the current limits on $|k^{(3)}_{(V),00}|$. From measurements of astrophysical birefringence, the limit 
$|k^{(3)}_{(V),00}|<2\cdot10^{-42}$ GeV was found in \cite{Kostelecky:2009zp, Carroll:1989vb}, and using Schumann resonances, the limit $|k^{(3)}_{(V),00}|<1.4\cdot10^{-20} \text{ GeV}$ was obtained in~\cite{Mewes:2008gh}. Plugging in the numbers, we find $\sqrt{\gamma}<10/3$ and $\sqrt{\gamma}<2.33 \cdot 10^{22}$ respectively, which is significantly weaker than the CMB polarization result.

Since a non-zero $\beta$ implies spacetime-symmetry breaking, we must consider the possibility that this is a spurious signal arising from some unknown systematic, the most likely culprit being the instrument miscalibration angle $\alpha$. For the {\it Planck} mission, the combination $\beta+\alpha$ has been reported to be $0.31^\circ\pm0.05^\circ$ \cite{Planck:2016soo}, where the error bars represent the estimated statistical error. To this we add a systematic error of $\pm0.28^\circ$ \cite{Planck:2016soo}, after which the result is consistent with zero birefringence angle, since $\alpha$ and $\beta$ are degenerate; however, in \cite{Minami:2020odp}, the authors employed a simultaneous determination of $\alpha$ and $\beta$, which has been shown to be robust across frequency bands and foreground contamination. Using this method, the result $\beta = 0.35^\circ\pm0.14^\circ$ has been achieved whilst eliminating the systematic uncertainty of $0.28^\circ$, which is what we translate to improve the bound on $|k^{(3)}_{(V),00}|$. It is a non-zero bound at $|k^{(3)}_{(V),00}| = \left(7.32\pm2.94\right)\cdot 10^{-45} \text{ GeV}$, which is the most stringent constraint to date. A possible way to tighten this bound further may be to use Stokes vector rotation, as was done in \cite{Kostelecky:2008be}.

Apart from a Chern-Simons type $F\widetilde{F}$ term, several mechanisms in conventional physics exist which may produce a non-zero $\beta$, the main two being Faraday rotation and foreground polarization: Faraday rotation produces a frequency-dependent birefringence angle $\beta(\nu)\propto \nu^{2}$, which was ruled out in \cite{Eskilt:2022cff,Eskilt:2022wav}. Foreground polarization produced by dust in the galactic plane do impact the results to some degree; in \cite{Eskilt:2022cff}, the authors mask out the galactic plane with a sky coverage of $f_{\rm sky}=0.62$, and still obtain a non-zero $\beta$ at $99.5\%$ confidence level \cite{Eskilt:2022cff}. In a different analysis, the authors of \cite{Diego-Palazuelos:2022dsq} finds a less statistically significant (although still non-zero) $\beta$. Recently, a study employing the Standard Model Effective-Field Theory (SMEFT) determined that the observed cosmic birefringence angle cannot be generated by a Standard Model operator at the energy scales relevant for the CMB \cite{Nakai:2023zdr}.

From the point of view of spacetime-symmetry breaking, there is a third avenue for signal pollution: symmetry breaking from other sectors. In this Letter, we have considered spacetime-symmetry breaking in the photon sector only, but it should be noted that certain types of symmetry breaking in the gravitational sector could in principle pollute this constraint. For example, considering a simple case of explicit breaking, the Hubble function for a flat Universe can be written as \cite{ONeal-Ault:2020ebv,Nilsson:2022mzq}
\begin{equation*}
    E(z)^2 =\Omega_m^0 (1+z)^{3}+\Omega_r^0 (1+z)^{4 x_r}+\Omega_\Lambda^0 (1+z)^{x_\Lambda},
\end{equation*}
where the exponents $x_r$ and $x_\Lambda$ are constants arising from the symmetry breaking. This modified measure would alter the conformal-time integral (\ref{eq:etaint}), and is an example of countershading of symmetry violations, discussed in \cite{Bailey:2016ezm,Kostelecky:2008in}.
Even in the case of standard cosmological background evolution, the density parameters $\Omega_X^0$ have error bars at the percent level, adding to the overall uncertainty, which we have not taken into account in this work. 

\begin{acknowledgments}
This work was supported by CNES. NAN is grateful for discussions with Quentin G. Bailey and Eiichiro Komatsu, and for support by PSL/Observatoire de Paris.
\end{acknowledgments}

\bibliography{sample}% Produces the bibliography via BibTeX.

\end{document}